\documentclass{elsart}
\usepackage{epsfig}
\usepackage{amssymb} 
\begin{document}
\newcommand{\gk}{\vec{\gamma}\vec{k}}
\newcommand{\gE}{\gamma_0 E_k}
\newcommand{\ppl}{\vec{p}}
\newcommand{\bcm}{\vec{b}^{\star}}
\newcommand{\becm}{\vec{\beta}^{\star}}
\newcommand{\bepl}{\vec{\beta}}
\newcommand{\rcm}{\vec{r}^{\star}}
\newcommand{\rpl}{\vec{r}}
\newcommand{\A}{{$\mathcal A$}}
\newcommand{\wpk}{ \omega_{p-k}}
\newcommand{\Journal}[4]{ #1 {\bf #2} (#4) #3}
\newcommand{\NPA}{Nucl.\ Phys.\ A}
\newcommand{\PLB}{Phys.\ Lett.\ B}
\newcommand{\PRC}{Phys.\ Rev.\ C}
\newcommand{\ZPC}{Z.\ Phys.\ C}
\newcommand{\be}{\begin{equation}}
\newcommand{\ee}{\end{equation}}
\newcommand{\bea}{\begin{eqnarray}}
\newcommand{\eea}{\end{eqnarray}}
\begin{frontmatter}

\title{Isospin Dynamics in Peripheral
Heavy Ion Collisions at Fermi Energies}

\author[catania,ctdip,florence]{J. Rizzo}
\author[catania]{M. Colonna}
\author[catania,ctdip,bucha]{V. Baran}
\author[catania,ctdip]{M. Di Toro}
\author[muenchen,catania]{H.H. Wolter}
\author[smith]{M. Zielinska-Pfabe}

\address[catania]{Laboratori Nazionali del Sud INFN, I-95123 Catania, Italy} 
\address[ctdip]{Physics and Astronomy Dept., University of Catania}
\address[bucha]{University of Bucharest and NIPNE-HH, Bucharest, Romania}
\address[florence]{Dip. di Fisica, University of Florence, Italy}
\address[muenchen]{Fak. f. Physik, Universit\"at M\"unchen,
D-85748 Garching, Germany}
\address[smith]{Smith College, Northampton, Mass. USA}
\address{e-mail: colonna@lns.infn.it}
\begin{abstract}


We present a detailed study of isospin dynamics in peripheral
collisions
at Fermi energies. We consider symmetric and mixed collisions of $^{124,112}Sn$
isotopes at 35 and $50 AMeV$
to study the isospin transport between the different reaction components
(residues, gas
and possibly intermediate mass fragments) and, in particular, 
the charge equilibration in the mixed system. We evaluate the effects of
drift terms due to asymmetry and density gradients, which are 
directly related to the poorly known value and slope of the symmetry energy 
below 
saturation density. We verify the importance of an isoscalar  momentum 
dependence of the mean field, which is found to influence the isospin transport
since it changes the reaction times. We finally suggest two observables
particularly sensitive to the isovector part of the nuclear equation-of-state:
the correlation between isospin equilibration and kinetic energy loss for
binary events, and the isospin content of the produced mid-rapidity fragments
for neck fragmentation events. 
\end{abstract}
%
\end{frontmatter}

\date{\today}


\section{Introduction}

 There has been much interest in recent years
 in
the determination of the nuclear symmetry energy as a function of density, 
which is
of great importance for nuclear structure as well as for astrophysical
 processes. While
 there has been much experimental and theoretical work in this direction in 
the last
years, the symmetry energy must still be considered as uncertain. Observables, 
which
are sensitive to the isovector density dependence of the 
Equation-of-State (iso-EoS) 
and testable
experimentally, are still under investigation
\cite{colonna98,BaoINTJE7,Isospin01,baranPR,WCI_betty}.

The symmetry energy $E_{sym}$ appears in the energy density
\be
\epsilon(\rho,\rho_3) \equiv \epsilon(\rho)+\rho E_{sym}(\rho)~(\rho_3/\rho)^2
 + O(\rho_3/\rho)^4 +....,
\label{esym} 
\ee
 expressed in terms of total ($\rho=\rho_p+\rho_n$)
 and isospin ($\rho_3=\rho_n-\rho_p$) densities. The symmetry term  has  a
kinetic contribution directly from basic Pauli correlations and a potential
part from the highly controversial isospin dependence,  as a function of 
the total nucleon density,
 of the  mean field 
\cite{baranPR}. Both at sub-saturation and supra-saturation
densities, predictions based on the existing many-body techniques diverge 
rather widely,  see ref.  \cite{fuchswci}.  However,  the dominant quadratic
dependence
 on  the asymmetry parameter $\beta=(\rho_3/\rho)=(N-Z)/A$  in  Eq.(\ref{esym})
is well supported by the phenomenology (e.g. the mass formula) as well as by 
all microscopic nuclear 
many-body calculations \cite{Isospin01}.
We  recall  that the knowledge of the 
 EoS  of asymmetric matter is very important at low densities ( e.g.  
neutron skins,
 pigmy resonances, nuclear structure at the drip lines, neutron distillation 
in fragmentation,
 neutron star formation and crust) as well as at high densities ( e.g.
 neutron star mass-radius relation, cooling, hybrid structure, transition
to a deconfined phase, formation of black holes).

It is attractive  to take advantage of new opportunities in
 experiments (availability of very asymmetric radioactive beams, 
improved methods of measuring event-by-event correlations) and
theory (development of reliable microscopic transport descriptions for heavy
ion collision (HIC)) to obtain 
results  which constrain  the existing effective interaction
models. In this paper we will discuss dissipative collisions in the range of 
    Fermi energies,
 which  will yield  information on the symmetry term around  and below  normal 
density.  Here we  focus our attention on the charge equilibration dynamics
in peripheral collisions where
we expect to see symmetry energy effects on the isospin transport. The 
interesting feature  at  Fermi energies is the onset of collective flows
due to compression and expansion of the interacting nuclear matter. The 
isospin transport takes place in regions with   density and asymmetry 
variations
and thus 
we expect to have contributions  to the isospin current from charge and mass
drift mechanisms. It was shown \cite{isotr05}, that these are determined by 
the value and the
 density-gradient  of the symmetry energy, respectively \cite{isotr05}, 
which are then studied here at
subsaturation densities.  Fast  (pre-equilibrium) nucleon emission, related
   to the symmetry energy
via the different symmetry potentials seen by neutrons and protons, also
influences the isospin content of the Projectile-like and Target-like
 residues.  Finally
in the Fermi  energy  range we have a substantial Intermediate Mass 
Fragment (IMF) production even in
non-central collisions and this will also affect the isospin dynamics.
We will carefully look at all these dynamical effects, trying to select
observables  which are  sensitive to the density dependence of the symmetry 
term.

Isospin equilibration at intermediate energies and its
relation to the symmetry
energy  has, in fact,  attracted much attention in recent years in  experiment 
as well as in theory \cite{isotr05,tsang92,soul04,shiPRC68,BALi}.
We present here a detailed analysis of the various iso-transport mechanisms
 in connection to properties of the effective in-medium interactions. We try
to disentangle effects due to the general reaction dynamics, ruled by
isoscalar terms of the interaction, from genuine isovector contributions. 
This will allow us to suggest new observables particularly sensitive
to the symmetry term of the EoS:
\begin{enumerate}
\item{Correlation between the isospin equilibration of the reaction
partners and the total kinetic energy loss in binary events;}
\item{The isospin content of IMF in break-up (ternary) events.}
\end{enumerate}
In Sect.2 we introduce  our transport-theoretical  approach to the reaction
dynamics  and the choice of the effective interactions investigated. In 
Sect.3 we show results for binary events in $Sn+Sn$ peripheral collisions at
50 AMeV, in Sect.4 we discuss  neck  fragmentation events,  and finally in
Sect.5 we  present a summary and some perspectives. 

\section{Theoretical  Description of the Collision Dynamics }

\subsection{The  Transport Model }

We perform {\it ab initio} collision simulations using
the microscopic Stochastic Mean Field (SMF) model.
It is based on mean field
 transport theory  with correlations included via hard nucleon-nucleon (NN)
collisions and  with inclusion of stochastic forces acting on the mean
phase-space trajectory
\cite{baranPR,guarneraPLB373,colonnaNPA642,fabbri04,chomazPR}.  Stochasticity 
is
essential in
order   to allow the growth of dynamical
instabilities with fragment production, as well as to obtain physical widths
of distributions of observables.  Moreover  it  will allow to perform
 event-by-event  correlation studies of great importance for  the very
complex reaction dynamics in this energy regime.

The transport equation for the phase space distribution function, with the
Pauli blocking consistently evaluated, is integrated following a
 representation in terms of test particles of finite widths
\cite{guarneraPLB373,colonnaNPA642}. A detailed description of the procedure
is given in ref. \cite{baranPR}. Our code \cite{Alfio}
has been extended  by the introduction of momentum dependent mean fields
 (see next subsection), which are 
 rather important in this energy range.  It has also  been possible to
improve the numerical accuracy while even reducing the computing times
\cite{rizzoj_th}.

A parametrization of free nucleon-nucleon cross sections is used, with
isospin, energy and angular dependence  \cite{LiMachl94}. Low energy
NN collisions, mostly
forbidden
because of the Pauli blocking,  have  large cross sections and
 could  induce spurious effects in  the  presence of some numerical
inefficiency in the blocking procedure, due to the discretization of the
phase space.
In order to  avoid such problems  a cutoff value $\sigma_ {cut}=50mb$ is used
in our calculations.  A parallel ensemble method is employed in the 
implementation of the collision term. 

For   discussions of isospin dynamics  in this energy
regime it is essential  to have a reliable
procedure for fragment recognition, i.e. to  identify  the  ``gas'' 
(emitted nucleons and light clusters)
and the  ``liquid''  (fragments, residues) phases.
 Fragments are produced
via instabilities in the expanding dilute matter and as residues of the
initial colliding ions (in the case of non-central collisions). 
In our transport simulations   the fragment definition  is performed by means 
of density
cuts ; more precisely, a criterion of $\rho>\rho_{cut}$ (with $\rho_{cut}$ 
between $\rho/6$ and $\rho/10$) for the 
liquid,
and correspondingly for the
 gas, is used. This procedure is applied in an event by event analysis at the
``freeze-out'', i.e. when the resulting fragments are well separated in space
and interacting only via Coulomb forces. We have checked that such ``final''
 gas/liquid isospin properties are not depending on the choice of density cuts 
and freeze-out times, see Sect.3.2.
Nucleons can be emitted during the early stages of the reaction
due to hard NN collisions as well as to repulsive     potentials, during
the fragment formation due to  the isospin  distillation effect,  and
finally from sequential decay
of excited primary products (fragments/residues).
The isospin is used in all these dynamical paths both as a tracer
of the reaction mechanism,
as well as an observable of interest with respect to the iso-EoS.

We follow the reaction dynamics only up to a ``freeze-out'' time.
Fragments and residues
will still be highly excited and will undergo sequential decays that can modify
the original isospin information related to the nuclear dynamics.
However, we
will analyze mainly quantities  which are expected to change little
due to the late evaporative
emission. 
 ``Imbalance'' and ``Double'' Ratios have been suggested to 
that purpose \cite{tsang92,famiano,MCcentral}, 
and will be discussed below. 

\subsection{Specification of  Interactions }
We adopt 
a  generalized form of effective interaction, which can be easily
reduced to Skyrme-like forces, with momentum dependent terms  also  in the
isovector channel  \cite{rizzoj_th,rizzoPRC72,ditoroAIP05}.
The general structure of this  isoscalar and isovector Momentum Dependent (MD) 
 effective interaction is
derived via an asymmetric extension of the Gale-Bertsch-DasGupta  (GBD)  force
 \cite{GBD,GalePRC41,GrecoPRC59,BaoNPA735},  which  corresponds to a Yukawian
 non-locality.\\
The energy density   is  parametrized
as follows:
 \begin{equation}
\varepsilon=\varepsilon_{kin}+\varepsilon_A(A',A'')
+\varepsilon_B(B',B'')+\varepsilon_C(C',C'') \quad,
\label{edensmd1}
\end{equation}
where $\varepsilon_{kin}$ is the usual kinetic energy density and
  the potential terms are  
\begin{eqnarray}
&&\varepsilon_A(A',A'')=(A'+A''\beta^2)\frac{\varrho^2}{\varrho_0}
 \nonumber \\
&&\varepsilon_B(B',B'')=(B'+B''\beta^2)\left ( \frac{\varrho}{\varrho_0}
\right )^{\sigma}\varrho
 \nonumber \\
&&\varepsilon_C(C',C'')=C'(\mathcal{I}_{NN}+\mathcal{I}_{PP})
+C''\mathcal{I}_{NP}
\label{edensmd2}
\end{eqnarray}
The variable $\beta=(N-Z)/(A)$ defines the isospin content  or asymmetry  of 
the system,
given the number of neutrons $(N)$, protons $(Z)$, and the total mass $A=N+Z$;
$\varrho$ is the nuclear matter density ($\varrho_0$ is the saturation value).
The momentum dependence is contained in the $\mathcal{I}_{\tau \tau'}$
terms, which are integrals of the form   
\begin{equation}
\mathcal{I}_{\tau \tau'}=\int d \vec{p} \; d \vec{p}\,'
f_{\tau}(\vec{r},\vec{p})
 f_{\tau'}(\vec{r},\vec{p}\,') g(\vec{p},\vec{p}\,') ~,
\label{momint}
\end{equation}
  with $\tau={P,N}$, for protons and neutrons.
Here  $f_{\tau}(\vec{r},\vec{p})$ are the nucleon phase space distributions
 for protons and neutrons
and the function  $g(\vec{p},\vec{p}\,') \equiv g[(\vec{p}-\vec{p}\,')^2]$
 determines the type of momentum dependence.
 A Skyrme-like momentum dependence is obtained  when we use the simple 
quadratic
form $g(\vec{p},\vec{p}\,') = (\vec{p}-\vec{p}\,')^2$.
A more general momentum dependence, in better agreement
with phenomenological optical potentials,
can be introduced by the function \cite{GBD,GalePRC41,GrecoPRC59,BaoNPA735}
\begin{equation}
g(\vec{p},\vec{p}\,')=
\left[
1+ \left( \frac{\vec{p}-\vec{p}\,'}{\Lambda}\right)^2
\right]^{-1} . 
\label{momform}
\end{equation}
We remark that this form
is particularly suitable for $SMF$ simulations.

From the energy density one derives the mean field potentials as
$U_{\tau}(\vec{r},\vec{p})=\delta\varepsilon/\delta f_{\tau}$. Thus the above
energy density implies a momentum dependent mean field interaction.
The momentum dependence is isoscalar if the coefficients $C'$ and $C''$ are
identical and also it can have an isovector part, if they are different.
The isovector momentum dependence implies different effective masses
for protons and neutrons given as
$\frac{m^*_{\tau}}{m}=(1+\frac{m}{\hbar^2 p}\frac{\partial U_{\tau}}
{\partial p})^{-1}$.

We note that the used form of the Momentum Dependent effective interaction,
Eqs.(\ref{edensmd1} - \ref{momform}), is
equivalent to the $MDI$ force introduced in ref.\cite{BaoNPA735}. The choice 
of the parameters is different: having fixed the isoscalar properties of the
EoS ($soft$ symmetric matter $K_{NM}(\rho_0)=215 MeV$, with nucleon 
effective mass $m^*/m=0.67$), we are exploring the dynamical effects of a
different density dependence of the symmetry term, isovector part of the
EoS, see Eq.(\ref{esym}). 
In Table I we report the used parametrization.

\noindent

\begin{center}
{\bf Tab. I:}~ Parameters of the Momentum Dependent (MD) Interaction.
\noindent

\begin{tabular}{c|c|c|c|c}
Isoscalar   &      & Isovector & Asy-stiff & Asy-soft       \\
\hline
$A'$	 &$-55.626\,MeV$ & $A''$ & $-41.312\,MeV$ & $59.742\,MeV$	\\
\hline
$B'$	 &$63.013\,MeV$	& $B''$	 &  $29.826\,MeV$ & $-71.247\,MeV$	\\
\hline
$C'$	 &$-129.41\,MeV$ & $C''$ & $-1354.78\,MeV$ & $-1354.78\,MeV$	\\
\hline
$\sigma$ &$1.242$ &   &   &   		\\
\hline
$\Lambda$ & $2.106\,fm^{-1}$ &  &  &	\\
\hline
\end{tabular}
\end{center}

Here we want to test the sensitivity of isospin transport observables
to two essentially different behaviors of the symmetry energy around 
saturation:
{\it asy-soft} with a smoothly decreasing behavior below saturation, and
{\it asy-stiff} with instead a rapid decrease to lower densities
\cite{colonna98,baranPR}.
In Fig.1 we show the density dependence for these two typical 
choices. 
We remark that this is the total symmetry energy, with both potential 
and kinetic contributions. 

\begin{figure}
\unitlength1cm
\begin{center}
\epsfig{file=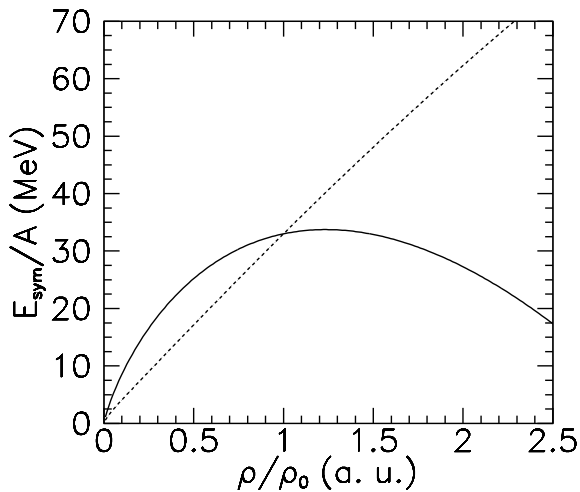,width=8.0cm}
\caption{Density dependence of the  symmetry energies  used in the simulations
presented here:  asy-soft (solid) and asy-stiff (dashed).}
\end{center}
\label{esymfig}
\end{figure}

 At Fermi
energies we have substantial  density variations during the collision.
In particular during the expansion phase the system  probes  more dilute
regions. Then we expect   important effects  on the isospin transport
 due to different subsaturation values (larger
for {\it asy-soft}) and to different slopes (larger for {\it asy-stiff}) 
of the symmetry 
energy. 
 We will see that
the  transport coefficients due to density and isospin gradients
(usually called ``drift'', resp. ``diffusion'' coefficients)  will be affected 
in a
different way
and the final result will be sensitive to the  choice of the two $E_{sym}$ 
shown in Fig.1. 

In the framework of our parametrization of the effective interaction
we can also easily adjust    the isospin momentum dependence  
to have the same density
dependence of the symmetry energy
{\it but  opposite neutron/proton  effective mass splittings}.
 The ordering of the neutron/proton effective masses, which is
also related to the slope of the Lane potential with energy,
is much debated in the literature \cite {baranPR,fuchswci}.
E.g. one could have $m^*_n>m^*_p$ as predicted by the
early Skyrme forces \cite{LyonNPA627} or the opposite as for the later
Skyrme-Lyon parametrizations
\cite{LyonNPA635}.   Thus one can,
separately from the density dependence,  study the corresponding dynamical 
effects
on nucleon emissions \cite{rizzoj_th,rizzoPRC72,ditoroAIP05}.

The effective mass splitting is increasing with density and
it is relevant for the dynamics mainly at high momenta. Indeed, it has been 
shown
 that the effect on nucleon emission and flows can be
observed only at high transverse momenta
 \cite{rizzoj_th,rizzoPRC72,ditoroAIP05}.
 At the energies investigated here it is, however, a secondary effect,
 as the transport phenomena are mainly driven by the value and the slope
 of the symmetry energy below saturation density. Thus, in order to
 simplify the presentation, we have fixed the isovector momentum dependence. 
The results presented here are obtained  with the  choice $m^*_n>m^*_p$.
We have actually  verified that the influence of the different mass splittings
on the present results on isospin
transport is very small. 

\subsection{Symmetry Potentials}

Many investigations have given definite evidence that the effective mean
fields are momentum dependent, i.e. have an important isoscalar momentum
dependence (see e.g. ref. \cite{DanLac02}).
Only in this way it was possible to explain the dependence with energy of the
nucleon flow in heavy collisions. We have then naturally included a momentum
dependence of the isoscalar interaction adjusted to these data, as discussed
before. However,
it is still of
interest to see the effects of such isoscalar momentum dependence on isospin
transport phenomena, since this has also been an issue in earlier calculations
\cite{BALi}, and has not been looked at systematically so far. Thus in the
following we will also show
results with  Momentum Independent (MI) interactions along with the Momentum
Dependent (MD) ones, of course with parameters adjusted to the same saturation 
properties and symmetry terms. 
Details about the MI interaction can be found in ref. \cite{baranPR}. 

We remark that the corresponding symmetry potentials, i.e. the effective field
variations seen by neutrons and protons in asymmetric matter, can be rather 
different
in the two frames, MI vs MD, and this could affect the isospin 
transport properties. So this point deserves some more attention.

\begin{figure}
\unitlength1cm
\begin{center}
\epsfig{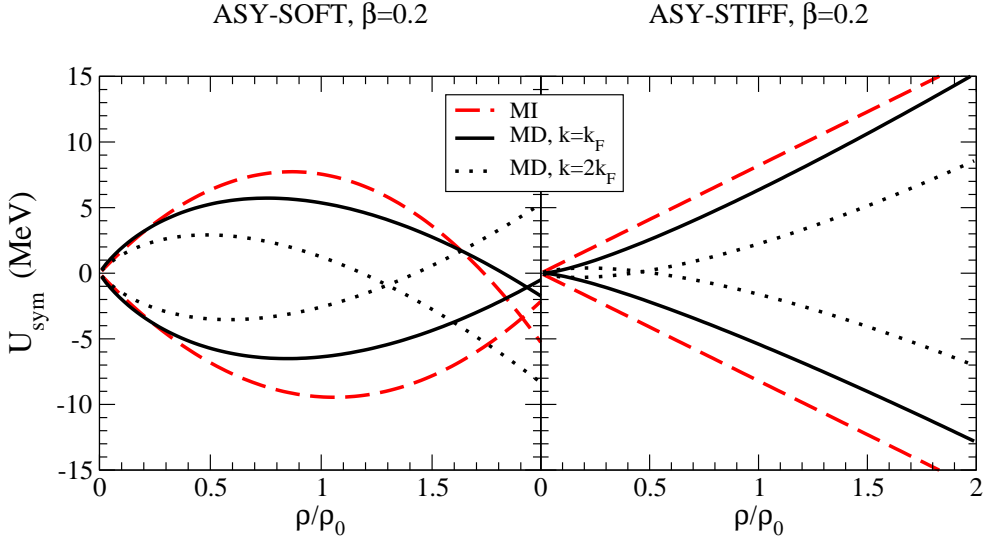}
\caption{Symmetry potentials for neutrons (upper curves) and protons 
(lower curves) in a matter with asymmetry $\beta=0.2$ ($^{124}Sn$-like).
Left panel: asy-soft choice; right panel: asy-stiff choice;
MI interaction: dashed lines, MD interaction:
solid ($k=k_F$) and dotted ($k=2k_F$) lines}
\end{center}
\label{potsym}
\end{figure}

Just to have an idea of the expected transport effects we plot
in Fig.2 the symmetry potentials for  
asymmetry matter with $\beta=0.2$, the mean asymmetry of a $^{124}Sn$ isotope.
We note that the momentum dependence, in fact, 
modifies the effect of the stiffness of the symmetry term 
on the nucleon potentials. E.g. comparing the {\it asy-soft} vs. 
{\it asy-stiff} choices 
 for neutrons and for MI interactions, we
see the general trend of a larger repulsion below
saturation and a larger attraction above $\rho_0$ (and opposite for protons).
It is also seen, that the effect can be different in the MD case, though 
depending strongly
on the nucleon momenta. In this respect we remark that, in particular below 
the saturation density, the differences become appreciable for nucleon
momenta around $2k_F$, not much present in the Fermi energy range of
interest here.

At variance a relevant overall effect of the Momentum Dependence is 
the reduction of the 
interaction times during the collision with an expected clear influence on the 
isospin diffusion.
An important consequence is that 
in order to extract information on the density dependence of the symmetry term
we are forced to select new observables related to the isospin equilibration
that should be not very sensitive to isoscalar properties of the effective
interactions, like the main momentum dependence. We will discuss in detail
this point in the next sections, suggesting very promising possibilities.

\section{General Features of the Isospin Transport in Peripheral Collisions}
\label{}

\begin{figure}
\unitlength1cm
\begin{center}
\epsfig{file=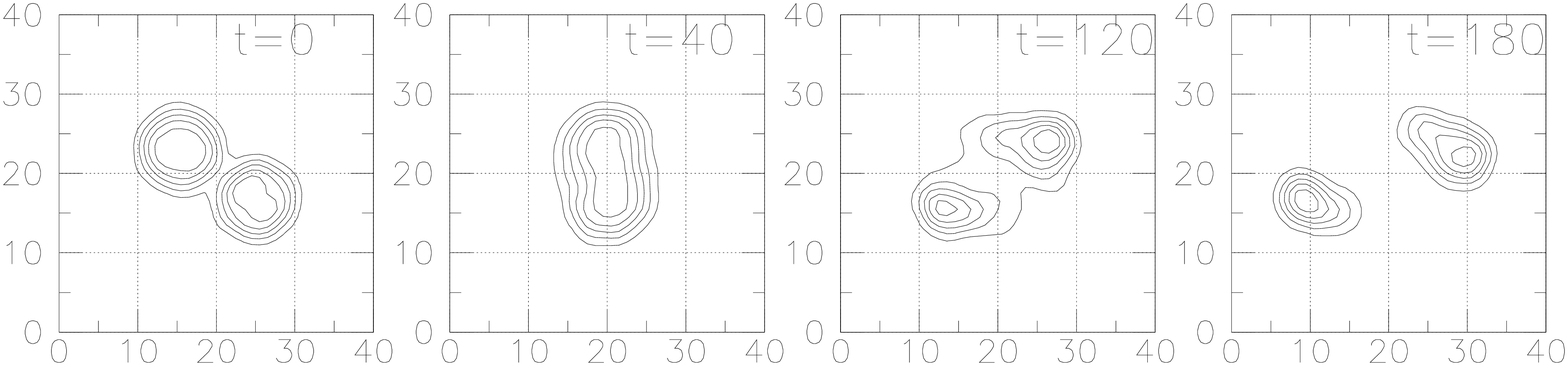,width=12.0cm}
\epsfig{file=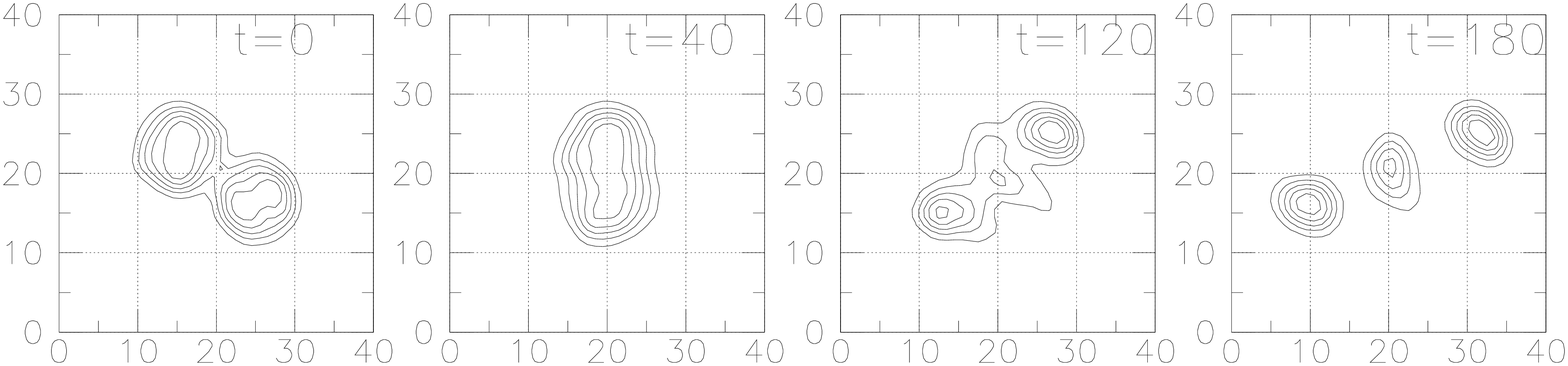,width=12.0cm}
\caption{Density contours of a binary (upper panel) and a ternary 
(lower panel) event
in the simulation of $^{124}Sn+^{124}Sn$ semiperipheral ($b=6fm$)
 collisions at $50AMeV$.
A momentum dependent interaction ($MD$) has been used.}
\end{center}
\label{binter}
\end{figure}

In this work we investigate peripheral collisions between similar systems with
different isospin, specifically collisions of different combinations 
of $Sn$ isotopes.
 In Fig.\ref{binter} density contours
of two typical examples of such events are shown, one in the upper panel
resulting in two
heavy
fragments  and nucleons or light clusters (binary event), and one in the
 bottom panel
 producing
a third fragment of intermediate mass from the neck region between the two 
heavy residues
(ternary event,  neck fragmentation ). 
 We will discuss binary events in Sect. 4, especially with respect
to isospin equilibration between the residues, and ternary events in
Sect. 5 with respect to the isospin content of the neck fragments. 

\subsection{Phases of the Reaction Mechanism}
The reaction mechanism can be divided into different phases, 
which will
 be important for the following discussion:

\begin{enumerate}
\item
{ Pre-equilibrium phase: the approach and early interaction phase is
characterized by an
emission of particles and light clusters (pre-equilibrium emission), which
contributes to
the ``gas'' phase. Of interest is the asymmetry of the gas, which has recently
been
discussed in detail in central collisions in similar systems, in particular, 
with respect
to ratios of produced particles and fragments \cite{famiano,MCcentral}.}
\item
{Transport phase: a neck of density below normal values develops between
the two
heavy residues, the evolution of which is driven by the motion of the 
spectators. During
this phase isospin is transferred to the neck due to the density difference 
between the
neck and the residues;  this effect  has been called isospin migration 
(isospin drift), which leads to a more
neutron-rich neck \cite{baranNPA703,baranNPA730}.  Moreover, in collision
systems with different asymmetry,
isospin is also   transported through neck
due to the $N/Z$ ``concentration'' difference, leading to an equilibration of
the isospin of the residues
 (projectile- and target-like fragments; PLF/TLF).
This has been called isospin diffusion.  In  asymmetric systems
there is
 a competition of isospin migration and diffusion, which contains information 
on the
characteristic time scales of the processes.}
\item
{Fragmentation into identified fragments  (primary fragments): as 
 in Fig.\ref{binter} we distinguish two types of events, binary events with 
two heavy residues
and the gas, and ``ternary'' events with an additional fragment of charge
$Z \ge 3$ and with mass between  the masses of the residues and the gas,
i.e. an intermediate mass fragment (IMF). 
Events with several
  IMF's  are very rare at these energies and impact parameters and will not
be considered.}
\item
{Evaporation stage: the primary fragments are considerably excited and
deexcite on the
 way to the detector, also changing their asymmetry. This part is not
included in our
 dynamical model, but could rather be accessed with statistical evaporation
codes. This
is not done here, because we try to use observables, which
 are expected not to change substantially in the  
secondary evaporation. It should be kept in mind, however, that the
 experimental  ``gas''
also contains contributions from the evaporation stage. }
\end{enumerate}

\subsection{Freeze-out Times and Density Cuts}

In the next sections we will present a detailed study of the isospin structure
of the reaction products analyzed at the freeze-out times, i.e. of the 
``primary'' nuclear systems (nucleons, clusters and fragments) emerging 
from the reaction region. Here we would like to show that the extracted isospin
 information is reliable, i.e. not much dependent on not well defined 
parameters, like the choice of the freeze-out times and of the density cuts
needed to separate the liquid and gas phases.

In  Fig.4 we show the time evolution of the ``gas'' 
asymmetry averaged over 200 event simulations of the 
$^{124}Sn+^{124}Sn$ semiperipheral ($b=6fm$)
 collisions at $50AMeV$, with the two choices of the symmetry stiffness
 (MD interactions). From the density contour plots of Fig.\ref{binter} 
we see that the freeze-out
time should be between 150 and 200 fm/c. The flat behaviour of the 
gas asymmetry after 150 fm/c clearly indicates that this quantity will not 
be much sensitive to a freeze-out choice in that interval. Of course the 
same will happen to the liquid phase.
The calculation has been performed using a $\rho_0/6$ (left panel)
and a $\rho_0/10$ (right panel) density cut for the 
gas phase. We can see that almost identical results can be obtained.
An interesting point to note is the larger final gas asymmetry with the 
{\it asy-soft} choice. This is a combined effect of the neutron distillation 
during the fragment formation (which gives the crossing at around 70 fm/c)
and of the larger repulsive potential below saturation, see 
Fig.2.

\begin{figure}
\unitlength1cm
\begin{center}
\epsfig{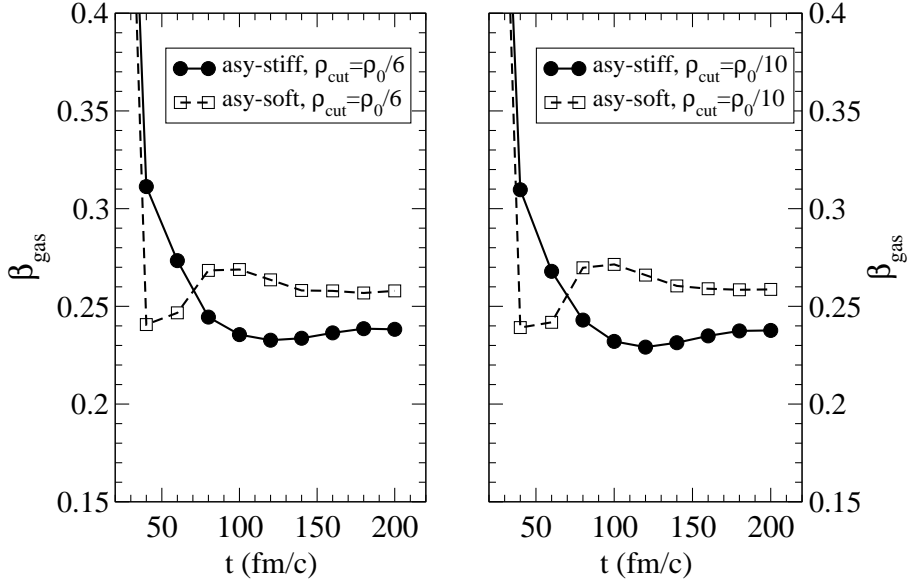}
\caption{Time evolution of the ``gas'' asymmetry
in the simulation of $^{124}Sn+^{124}Sn$ semiperipheral ($b=6fm$)
 collisions at $50AMeV$. Asy-soft: dashed lines. Asy-stiff: solid lines.
The panels correspond to different density cuts for the ``gas'' definition, 
see text.
A momentum dependent interaction ($MD$) has been used.
}
\end{center}
\label{bgascheck}
\end{figure}

\noindent
\begin{center}
{\bf Tab. II:}~$^{124}Sn+^{124}Sn$ collisions: Freeze-out times (fm/c)
 vs. beam energy, centrality and momentum dependence.
\noindent
\begin{tabular}{c|c|c|c|c|c|c|c}
35~AMeV  &  b~(fm)    & MD & MI & 50~AMeV &  b~(fm)  & MD & MI     \\
\hline
	 & $ 6 $  & $180$ & $200$ & 	& $6$  &  $140$ & $150$ \\
\hline
	 & $ 8 $ & $160$ &  $175$ &   &	 $8$  &  $120$ & $125$ \\
\hline
	 & $ 10 $ & $120$ & $125$ &  &	 $10$ & $100$ &  $100$ \\
\hline

\end{tabular}
\end{center}

Finally in Table II we give our quantitative 
estimates of the variation of the
freeze-out times with beam energy, centrality and momentum dependence of the
used effective interactions. The nuclear interaction times are different
for MD and MI interactions (shorter for MD) and this property of the 
isoscalar dynamics will certainly affect
the isospin diffusion. In order to extract the role of the isovector part of 
the interaction we have to single out observables directly related to the 
interaction times.

\subsection{Isospin Transport Coefficients}
As already pointed out,  in heavy ion collisions at Fermi energies
isospin transport
is due to the combined effect  of asymmetry and density gradients.
It can be discussed in a compact way by means of the chemical
potentials for protons and neutrons as a function of density $\rho$ and 
isospin asymmetry
$\beta$ \cite{isotr05}.  In fact,  the $p/n$ currents can be expressed as
\be
{\bf j}_{p/n} = D^{\rho}_{p/n}{\bf \nabla} \rho - D^{\beta}_{p/n}{\bf \nabla}
 \beta
\ee
with $D^{\rho}_{p/n}$ the  mass, and
 $D^{\beta}_{p/n}$  the  isospin transport coefficients, which are directly 
given
by the variation of $n,p$ chemical potentials with respect to density and
asymmetry,
 see ref. \cite{isotr05}. 
In the literature these are often referred to as
 the drift, resp. diffusion coefficients, and we will follow this convention
 here.  Of interest here is the differential current  of 
neutrons and protons (iso-vector current) which has a simple relation to the 
density
dependence of the symmetry energy. In fact, from the rather general parabolic
 form of
the symmetry term
in the energy density,   we obtain the important relation
 $\mu_n-\mu_p=4 \beta E_{sym}$ and derive 
 the drift and diffusion contributions to the isovector current
\bea
D^{\rho}_{n} - D^{\rho}_{p}  & \propto & 4 \beta \frac{\partial E_{sym}}
{\partial \rho} \, ,  \nonumber\\
 D^{\beta}_{n} - D^{\beta}_{p}  & \propto & 4 \rho E_{sym} \, .
 \label{drift}
\eea
Thus the isospin transport due to density gradients, i.e. isospin migration, 
depends on
the slope of the symmetry energy, or the symmetry pressure, while the 
transport due to
isospin     gradients, i.e. isospin diffusion, depends on the 
absolute value
 of the symmetry energy. In peripheral collisions discussed here, residues 
of about
normal density are in contact with the neck region of density below 
saturation density.
 In this region of density a stiff iso-EoS has a smaller value but a larger 
slope
compared to a soft iso-EoS. Correspondingly we expect opposite effects of 
these models
 on the migration and diffusion of isospin.

In our investigation we will see that the {\it asy-soft} choice 
appears to be more
effective for
isospin equilibration. This is the result of a dominant diffusion mechanism
for peripheral
 collisions, 
together with a stronger fast neutron emission, which also 
contributes to the $N/Z$ equilibration. With respect to the pre-equilibrium
emission we recall that in the {\it asy-soft} case neutrons see
a more repulsive
 symmetry potential around saturation, see Section 2.3.
On the other hand, the {\it asy-stiff} choice is more effective for isospin
migration, which will be seen to be important for the isospin content
of the neck fragments.

\subsection{The Imbalance  Ratio }
 We will discuss the asymmetries of the various parts of the reaction system
(gas,  PLF/TLF's, and in  the  case of ternary events,  IMF's ).
 In particular, we study  the
 so-called imbalance ratio (also called  Rami or transport ratio 
\cite{ramimb}), which is defined as
\be
R^x_{P,T} = \frac{2(x^M-x^{eq})}{(x^H-x^L)}~,
\label{imb_rat}
\ee
with $x^{eq}=\frac{1}{2}(x^H+x^L)$.
 Here, $x$ is an isospin sensitive quantity
that is to be investigated with respect to
equilibration.   In this work we consider primarily the asymmetry 
$\beta=(N-Z)/(N+Z)$,
but also other quantities, such as isoscaling coefficients, ratios of 
production of light
 fragments, etc, can be of interest \cite{WCI_betty}.  
The indices $H$ and $L$ refer to the symmetric reaction
between the
heavy  ($n$-rich) and the light ($n$-poor)  systems, while $M$ refers to the
mixed reaction.
$P,T$ denote the rapidity region, in which this quantity is measured, in
particular the
PLF and TLF rapidity regions. Clearly, this ratio is $\pm1$ in
the projectile
and target regions, respectively, for complete transparency, and oppositely
for complete
rebound, while it is zero for complete equilibration.

In a simple model we can show that the imbalance ratio mainly depends on two
quantities: the strength of the symmetry energy and the interaction
time between the two reaction partners.
Let us take, for instance, the asymmetry $\beta$ of the PLF (or TLF) as the
quantity $x$.
In the Fermi energy domain we can describe the charge equilibration dynamics 
as an overdamped dipole oscillation mode, see \cite{baranPRL87,kazim}.
Therefore,
 as a first order approximation, in the mixed reaction the charge asymmetry
parameter will show an exponential relaxation
towards
its complete equilibration value, $\beta_{eq} = (\beta_H + \beta_L)/2$ as
\begin{equation}
\label{dif_new}
\beta^M_{P,T} = \beta^{eq} + (\beta^{H,L} -  \beta^{eq})~e^{-t/\tau},
\end{equation}
where $t$ is the time elapsed while the reaction partners are interacting
(interaction time) and the damping $\tau$ is mainly connected to the strength 
of the symmetry energy. 
In fact, as seen in Eq.(\ref{drift}), isospin diffusivity is directly 
proportional
to the symmetry energy \cite{isotr05,shiPRC68}.
Inserting this expression into Eq.(\ref{imb_rat}), one obtains
$ R^{\beta}_{P,T} = \pm e^{-t/\tau}$ for the PLF and TLF regions, respectively.

From this simple result one sees that
the imbalance ratio
does not depend on the difference of asymmetries $(\beta_{H}- \beta_{L})$
of the systems considered,
at least at a first order level,
which is due essentially to the normalization to the difference
$(\beta_{H}- \beta_{L})$
in the definition of $R$ in Eq.(\ref{imb_rat}).
Hence the imbalance ratio can be considered as a good observable to 
trace back the strength
of the symmetry energy from the reaction dynamics. 
We will show that the
effect of different
iso-EoS choices on the final imbalance ratios are measurable, in
particular, when
correlated to the total energy loss of the dissipative collision, which sets 
the time-scale of the process.

\section{Binary Reactions}

In this section we   discuss binary reaction events, and investigate the
asymmetries of the various reaction components, i.e. of the gas,  the residues,
and the exchanged particles, as well as the imbalance parameter derived
from  these  via
Eq.(\ref{imb_rat}). We will consider these quantities for the different
isovector-EoS's,
 and for momentum-dependent (MD) and momentum-independent (MI) 
parametrizations of the
isoscalar part, see Sect. 2.2. 
The aim of the analysis is also to show that effects due to the 
Momentum-Dependent
term of the nuclear interaction are relevant for the reaction mechanism
at Fermi energies, and consequently for the isospin equilibration.
Here we  discuss these quantities as a
function of the incident
 energy
and impact parameter, in the Sect.4.3  we will introduce the more
significant correlation to the kinetic energy loss.

\begin{figure} [h]
\begin{center}
\unitlength1cm
\epsfig{file=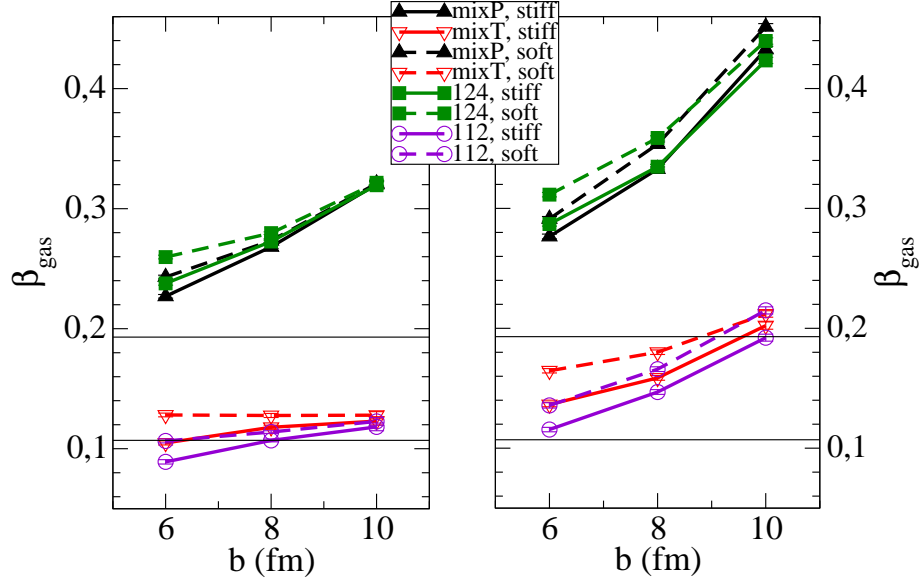,width=10.0cm,angle=-90}
\caption{Asymmetries of the gas in $Sn+Sn$ collisions at incident energy 
of $E=50  
AMeV$ for MD (left) and MI (right) interactions for  a  stiff (solid lines)
and soft 
(dashed lines) iso-EoS. The symbols differentiate the different reaction 
systems: 
mixed PLF (full triangle up), mixed TLF (empty triangle down), symmetric heavy 
(full squares), symmetric light (empty,  circles ). Thin horizontal lines 
denote the 
initial asymmetries of the heavy and light system, respectively.
}
\end{center}
\label{asymm_gas}
\end{figure}

\subsection{Asymmetries of  Reaction Components }

We first investigate the behaviour of the   asymmetries $\beta$ of the 
reaction components  for the different $Sn+Sn$ reactions for the impact 
parameters 
$b=6,8,$ and $10 fm$ and for two incident energies. We  present  our 
results for an 
incident energy of $E_{lab}=50 AMeV$ for the gas phase in 
Fig.5, for 
the residues in Fig.6, and for the exchanged particles in 
Fig.7. These figures are arranged in an  analogous  way: 
the results 
with the momentum dependent interaction (MD) are in the left panel, those 
with the 
momentum-independent one (MI) on the right.
 Results for the stiff and soft iso-EoS are given by solid and dashed lines, 
respectively. The symbols distinguish the different systems: mixed $HL$ 
reaction PLF 
(triangle up, full), TLF (triangle down, empty), symmetric heavy $HH$ 
(squares, full) and light $LL$ ( circles, empty) systems.   Thus, also
for the gas  we have
separated the  emitted  particles according to their origin from the
projectile (P) or target
(T). 
The  asymmetries
of the  exchanged particles are
only given for the mixed system. The initial asymmetries of $^{124}Sn$ and
$^{112}Sn$ 
are indicated as horizontal lines.

\begin{figure} [h]
\unitlength1cm
\begin{center}
\epsfig{file=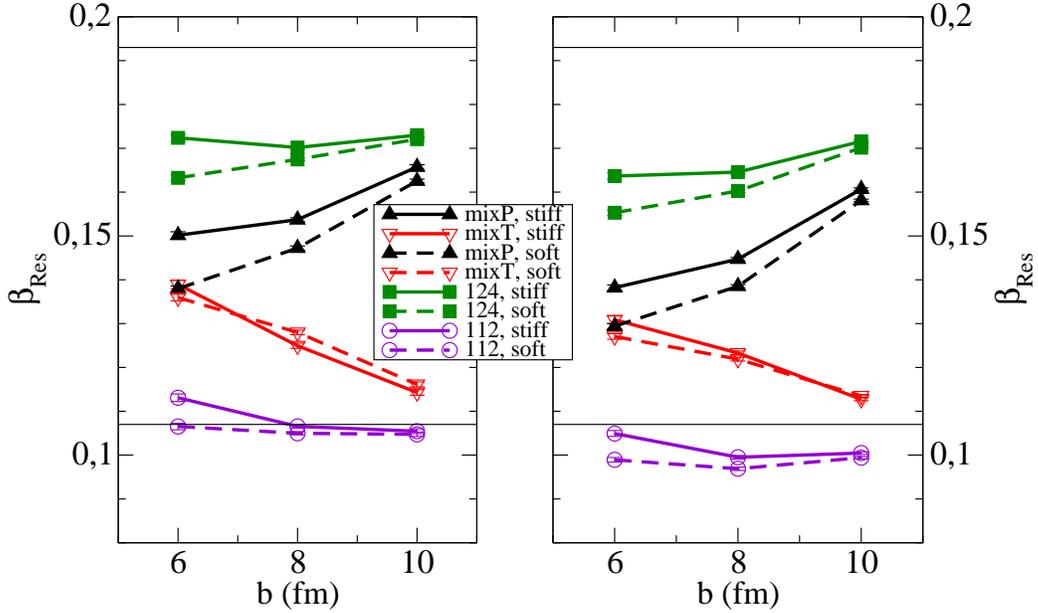,width=10.0cm,angle=-90}
\caption{Asymmetries of the residues in $Sn+Sn$ collisions at incident 
energy of 
$E=50 AMeV$ for MD (left) and MI (right) interactions. The arrangement
 of the figure 
and the meaning of the lines and symbols is as in Fig.5.
}
\end{center}
\label{asymm_res}
\end{figure}

\begin{figure} [h]
\unitlength1cm
\begin{center}
\epsfig{file=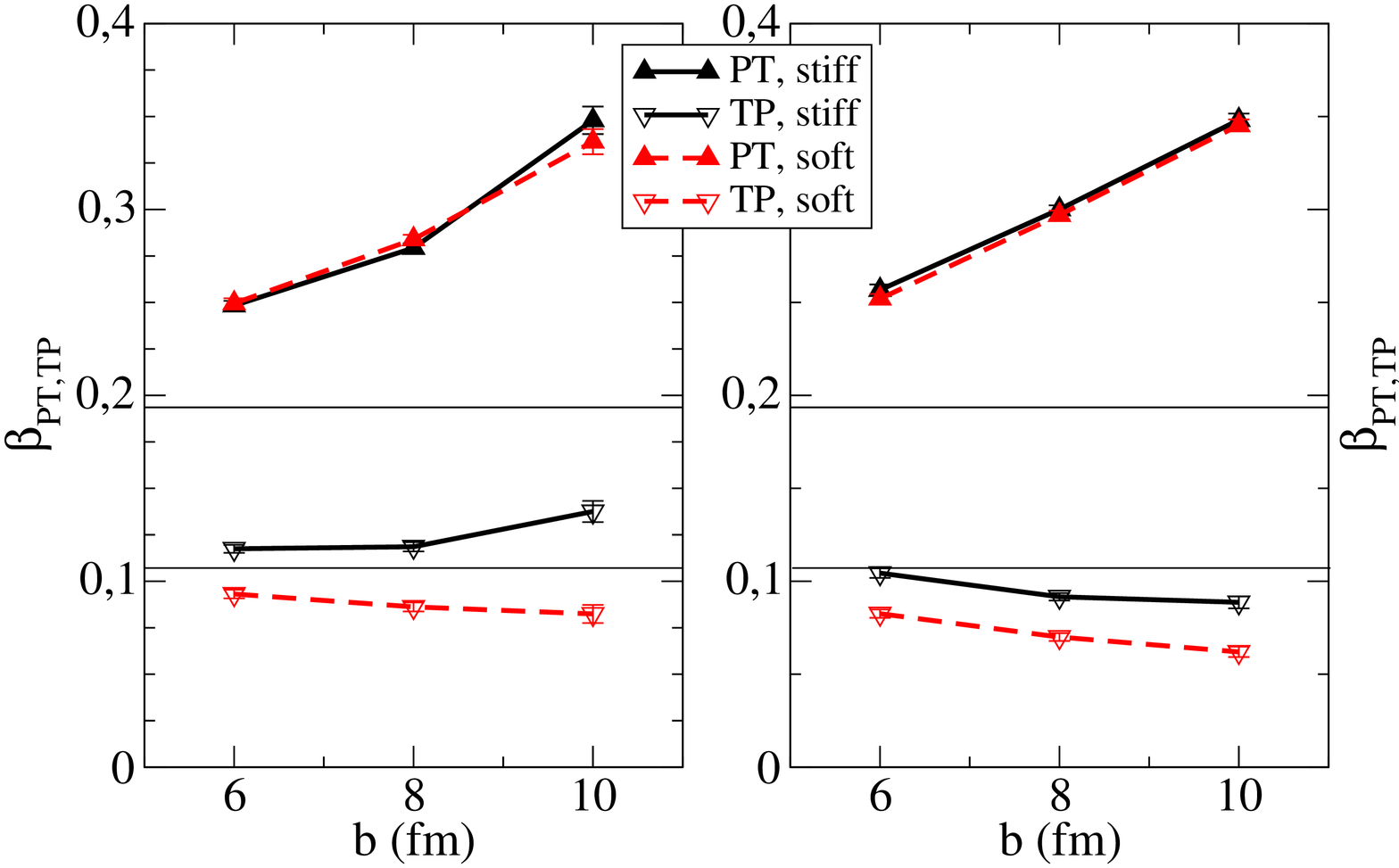,width=10.0cm,angle=-90}
\caption{Asymmetries of the exchanged particles in the collision 
$^{124}Sn+^{112}Sn$
 at incident energy of $E=50 AMeV$ for MD (left) and MI (right) interactions,
 and for
the stiff (solid lines) and soft (dashed lines) iso-EoS, respectively.  PT
 denote
 particles transferred from the projectile to the  target  (triangle up, full),
and  
vice versa for TP (triangle down, empty).
}
\end{center}
\label{asymm_exch}
\end{figure}

We first note the following general properties of these asymmetries, which 
will be 
valid for all cases discussed in the following.
Of course, all these asymmetries are
connected by
charge conservation. In absolute numbers the mass of the exchanged particles
 changes
from about 20 to 5 amu, and for the gas from about 40 to 20 amu, for $b = 6$
to $10 fm$,
 respectively.
\begin{itemize}
\item
{ The gas  (Fig. 5)  is generally more $n$-rich
(i.e. asymmetric) than the initial asymmetry of the corresponding nucleus;
always for
the $n$-rich nucleus (P) but mostly also for the $n$-poor (less $n$-rich) 
system (T). In a
nuclear medium with
neutron excess the interaction is more repulsive for neutrons, such that these
 are 
emitted preferentially.}
\item
  { The asymmetry of the residues  (Fig.6)  for the mixed 
$HL$ system
decreases for the   $n$-rich (PLF) and increases for the $n$-poor (TLF) 
partner with respect
to the initial asymmetries, as expected for isospin equilibration.
 It is interesting to look closer at the residue asymmetry 
$\beta_{Res}$ 
in the case of the $HH$ and $LL$ symmetric collisions. Here we cannot have
isospin 
transport and the only variations come from nucleon emissions.
The asymmetry decreases for the symmetric systems from the initial asymmetry,
less evident in the
$n$-poor system, in correspondence with
the asymmetry of the gas.
The change in the asymmetries of the residues of the mixed $HL$ systems is much
bigger than in the symmetric systems, indicating     the importance of the
transport contributions to the isospin current. This is also supported by the
dependence on the iso-EoS discussed below.}
\item
 { In the mixed $HL$ system neutrons are predominantly 
exchanged from the  $n$-rich to the $n$-poor system, i.e. the asymmetry of the
exchanged particles in this direction is much larger than the initial 
asymmetry of
the $n$-rich system. In the other direction the
asymmetry
is close to the initial asymmetry.}
 \item     
 {The impact parameter dependence clearly shows  that iso-EoS effects are more
relevant for more dissipative collisions, i.e. for smaller impact parameters 
and
thus for longer interaction times.  
This 
suggests the use  of an event selection in terms of the total kinetic 
energy loss 
in order to enhance the 
sensitivity to the symmetry energy, as discussed  in Sect.4.3. 
}
\end{itemize}

Next we note the differences when using the MD or MI  interactions  in the 
isoscalar sector. 
It is seen that the gas is generally more symmetric for a MD interaction. 
The MD 
interaction has a stronger isospin-blind repulsion, which emits protons and 
neutron 
equally. Thus the mass of the emitted gas is about 5 to 10 units higher in
the MD case. 
This reduces the relative effect of the isovector interaction. Correspondingly,
 the 
residues are generally more symmetric for the MI relative to the MD case  since
more of the neutron excess is emitted into the gas.  In addition,
in the MD case the reaction proceeds more rapidly, because of the  greater 
isoscalar 
repulsion, thus reducing the interaction time and reducing the amount 
of isospin 
equilibration. This is seen clearly in the mixed systems.

With respect to different iso-EoS's we see that the gas is more asymmetric
for the {\it asy-soft} case, which is expected because of the higher 
sub-saturation symmetry energy and thus 
the larger 
neutron repulsion.  Consequently the residues are more symmetric for the
soft iso-EoS both for the symmetric and the mixed systems. 
Also, the change of the asymmetries of the residues is much larger for the
mixed ($HL$) than for the symmetric systems, indicating that
isospin equilibration will be mostly sensitive to the iso-EoS via the transport
contributions of the isospin current.  
In the exchanged particles we see almost no effect for the particles
transferred in the 
direction projectile to target, and an interesting difference in the other 
direction.
 This is due to the fact that, in the direction from  projectile
to target, the larger isospin diffusion effects expected for the
{\it asy-soft} EoS are counterbalanced by the more neutron-rich emission to the
gas, which reduces the projectile asymmetry. On the other hand, diffusion
effects act in opposite direction for the $n$-poor source, with a
smaller transferred asymmetry from target to projectile in the {\it asy-soft}
case, also due to the larger neutron emission to the gas.
Thus in this direction we do expect a sensitivity to the iso-EoS,
as seen in Fig.7. 

 However,   we have to
note that generally the changes due to different isovector EoS's are not 
very large. 
As we will see  below, the imbalance parameter considerably increases  the
sensitivity,
since the transport contributions will be enhanced.  In fact, it
was shown in ref. \cite{isotr05}
that the transport ratios depend mostly on $(\beta_{PT}- \beta_{TP})$.
i.e. the asymmetry difference
of the exchanged nucleons, and less on pre-equilibrium emissions.

\begin{figure} [h]
\begin{center}
\vskip 1.5cm
\unitlength1cm
\epsfig{file=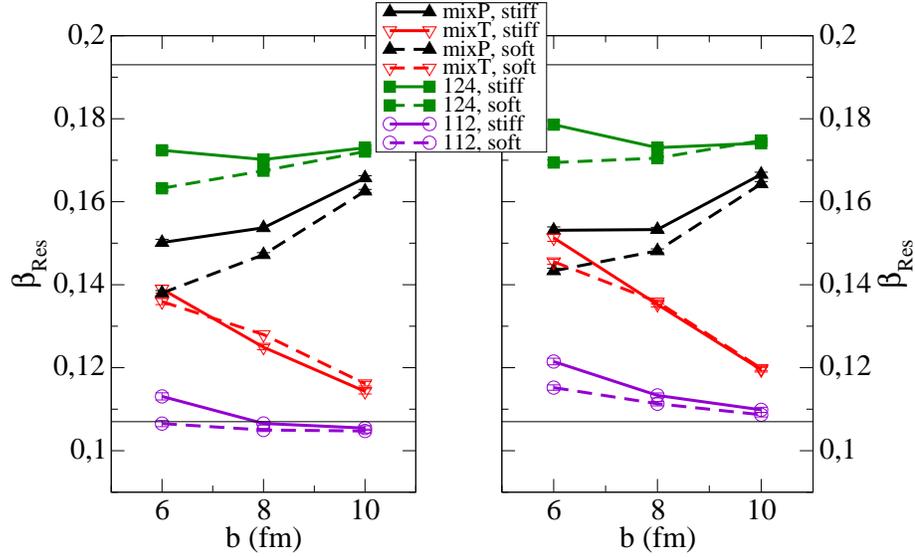,width=12.0cm}
\caption{Asymmetries of the residues in $Sn+Sn$ collisions at incident 
energies of 
$E=50 AMeV$ (left) and $35 AMeV$ (right) for MD interactions. The meaning 
of the lines 
and symbols is as in Fig.3.}
\end{center}
\label{asymm_res3550}
\end{figure}

Finally in Fig.8 we compare 
(for the case of MD interactions) 
the residue asymmetries for incident energies of 50 $AMeV$ 
(left panel, already shown in Fig.4) and 35 $AMeV$ (right panel). 
Competing effects are observed, since on one hand the emission to the 
continuum is 
lower at the lower energy, leading to less change of the initial asymmetry,
 and on the 
other hand, the interaction time is larger, leading to more equilibration 
in the mixed 
system.  

\begin{figure} [ht]
\begin{center}
\vskip 1.5cm
\unitlength1cm
\epsfig{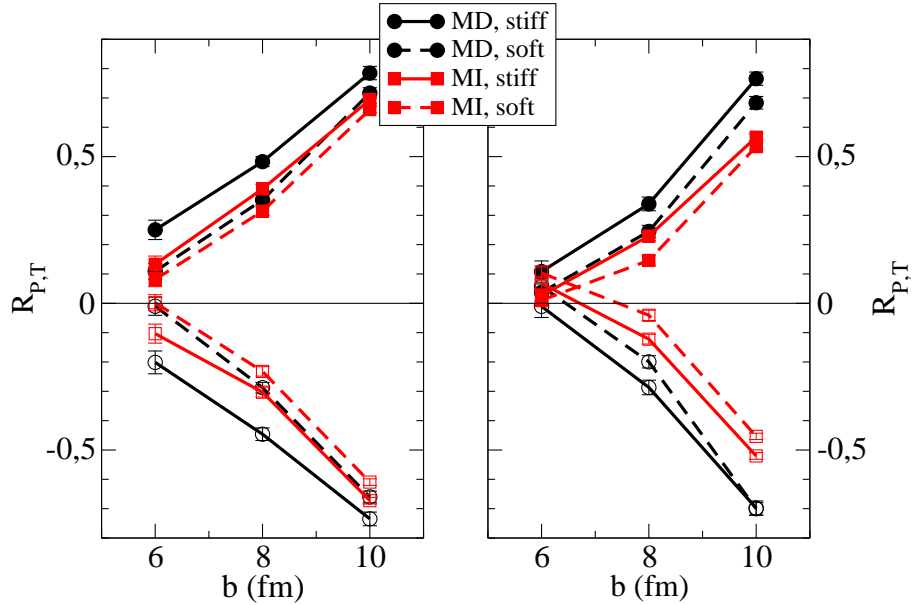}
\caption{ Imbalance ratios for $Sn + Sn$ collisions for incident energies
of 50 (left) 
and 35 $AMeV$ (right) as a function of the impact parameter. Signatures of 
the curves: 
iso-EoS stiff (solid lines), soft (dashed lines); MD interaction (circles),
 MI interaction (squares); projectile rapidity ( full symbols, upper curves ),
 target rapidity ( open symbols, lower curves ).}
\end{center}
\label{imb_Eb}
\end{figure}

\subsection{Isospin Imbalance Ratios}

From the asymmetries shown before one obtains the imbalance 
ratios
 according to Eq.(\ref{imb_rat}).  These measure  the amount
of isospin
equilibration in the mixed relative to the symmetric reaction systems,  
having  
lower absolute
 values for more equilibration.
 
 The results for the imbalance ratios are shown in Fig.9
for the incident energy of  50 $AMeV$ (left panel) and 35 $AMeV$  
(right panel), for MI/MD interactions and different iso-EoS choices.
Two very clear trends with respect to the amount of  equilibration  appear: the
 equilibration is 
larger ($R_P$ smaller) both, for lower energies  (right panel)  and for 
MI interactions
 (dashed curves).  In both cases this can be
 understood from the fact that the reaction is slower and thus the interaction
 time longer, 
leading to more equilibration; for the lower energy because of the lower 
speed, and for the MI
 interaction because of the less repulsive isoscalar mean field. In comparing
 the two 
iso-EoS's we see that the equilibration is larger for the soft iso-EoS, since 
the higher 
symmetry energy leads to a larger diffusion contribution to the 
isospin current, as 
discussed above. The first two effects 
(incident energy and MD interactions) are essentially kinematical, 
depending on the interaction 
time, while the last one depends on the iso-EoS, which is of interest here. 
Thus, in the next 
section we will propose a scheme to separate the kinematical from the isospin 
effects.

The results in Fig.9  can be compared to other results in the
literature. In our
previous work \cite{isotr05} we have used MI  interactions  and  
slightly different 
parametrizations
of the iso-EoS, with a somewhat softer symmetry term.  The results here 
agree well with the
previous ones within these variations. 

On the other hand, there has 
been extensive
discussion of isospin imbalance ratios 
 in several papers of B.A. Li and collaborators \cite{BALi}, 
in which a different family of 
iso-EoS's has been used, which is characterized by an  index $x$ 
(for the density 
dependence of the symmetry energy around saturation of the 
form $\rho^{\gamma}$),
together with
momentum-dependent interactions,  and with  free and medium-modified cross 
sections.
The case $x=-1$, 
corresponding to $\gamma \approx 1$, and with free cross sections, should 
be compared with our stiff 
MD parametrization ($\gamma=1$). However, the value given in
ref. \cite{BALi} for $b=6fm$ of
$R_P \approx 0.42$ does not agree well with our value of  $R_P \approx 0.25$ 
 in Fig.9.
 The reason could reside  in ingredients of the BUU
transport code used
(point test particles, surface effects) leading to a
different treatment, relative to the SMF simulations here, of the isoscalar
part of the nuclear interaction. This will influence, in turn, the 
interaction
time between the two reaction partners and the degree of dissipation reached
in the collision.

The existing imbalance ratio measurements for $Sn$ isotopes 
have been analyzed with an impact parameter selection performed in terms of 
charged particle multiplicities \cite{tsang92}. A value $R_P \approx 0.45$ 
is extracted in the impact parameter range of $b=6-7.3 fm$, 
which is also not far from our {\it asy-stiff} MD estimations ($\gamma=1$ or 
slightly stiffer). From Fig.9 (left panel) we see that in this impact 
parameter range the imbalance ratio $R_P$ is rapidly increasing
and so it is very important to assess the same centrality selection 
in the comparison between data and simulations.    
To overcome these problems and isolate isospin effects, 
one could study the imbalance ratio directly as a function of 
the interaction time
(or an observable directly related to it),
as we will do in the next section.

\subsection{Correlation with  Kinetic Energy Loss }

In   subsection 4.1 we have noted,  e.g. in Fig.6, that the residues  
appear more symmetric for the momentum-independent interaction.  
 Also, we observe in Fig.8  that at lower energy the longer 
equilibration time  is counterbalanced by the smaller pre-equilibrium 
emission. These 
trends are more 
clearly seen in Fig.9 for the imbalance ratio, which directly measures the 
amount
 of equilibration. In fact the latter is 
clearly less effective when
 either the incident 
energy is higher or momentum-dependent interactions are used.  As we 
have remarked,
both effects decrease 
the contact time of the two nuclei in the binary collision. Thus these 
observations point to 
the interaction time as the dominating influence on the amount of  
isospin equilibration, as 
was already discussed in ref. \cite{isotr05}.

On the other hand, longer interaction times should be correlated with larger 
dissipation and, in particular, with damped PLF and TLF velocities.
The dissipation, in turn, has been measured,  e.g.  in deep 
inelastic collisions,
 by the kinetic
energy loss. Thus in this subsection we will investigate the correlation 
between the imbalance 
parameter and the kinetic energy loss. We define the kinetic energy loss per 
particle as
\be
E_{loss}  =  E_{cm} - \frac{E_{kin}+E_{pot}^{Coul}}{A_{PLF}+A_{TLF}} \, , \,
E_{cm}  =  \frac{E_{lab}}{A_P} \frac{A_P A_T}{(A_P+A_T)^2} \, ,
\label{E_loss}
\ee
where $A_P, A_T, A_{PLF}, A_{TLF}$ are the masses of the initial projectile 
and target, and of 
the final projectile-like and target-like fragments, respectively. Here 
$E_{cm}$ is the initial 
energy, per nucleon,  available in the $cm$ system. $E_{kin}$ and 
$E_{pot}^{Coul}$ are 
the final kinetic 
energies of the fragments in the $cm$ system, and the Coulomb potential 
energy, respectively. 
We will use the relative energy loss  $E_{loss}/E_{CM}$  as a measure of 
 PLF/TLF velocities and thus
of the interaction time. The study of isospin equilibration as a function of
the heavy residue excitation energy (related to the kinetic energy loss  )
was also suggested in ref. \cite{soul04}.

\begin{figure} [ht]
\unitlength1cm
\begin{center}
\epsfig{file=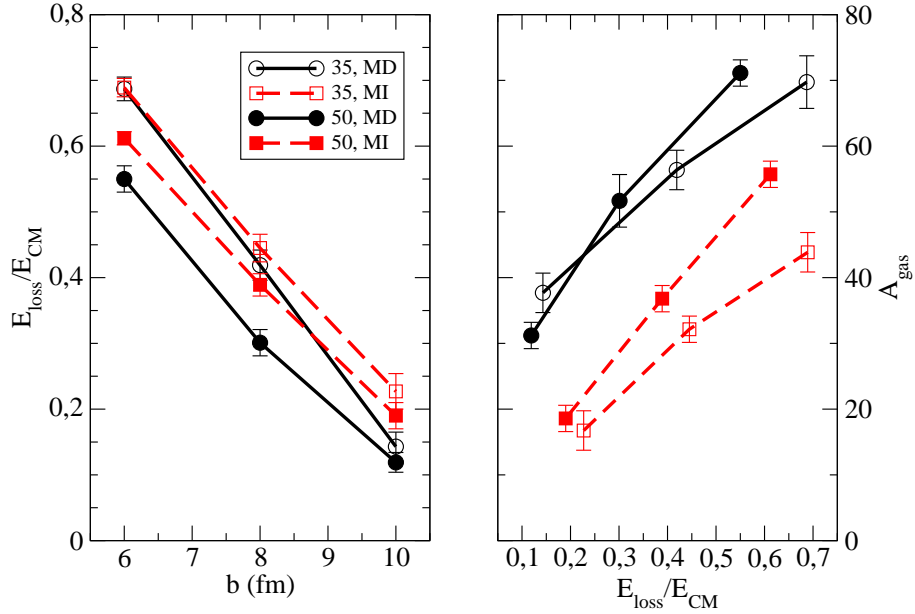,width=10.0cm,angle=-90}
\caption{ (left panel) Relative energy loss as a function of impact parameter; 
and (right panel) 
mass of particles emitted into the gas as a function of relative energy loss. 
Curves are shown for
 MD interactions (solid lines and circles) and MI interactions (dashed lines
 and squares), and for
 35 MeV (open symbols) and 50 MeV (full symbols) incident energy, respectively.
}
\end{center}
\label{eloss_b_gas}
\end{figure}

The relative energy loss is shown in Fig.10
(left panel) as a function of 
impact 
parameter for the two incident energies and MD, resp. MI interactions. It is 
seen that this 
quantity exhibits, generally, a very good correlation with impact parameter, 
except for the 
large impact parameters, where eventually all curves will converge to zero. 
As expected the 
energy loss at a given impact parameter is larger for lower energies and for 
MI interactions, 
because of the longer interaction times. 

However,
the energy loss consists not only in internal excitation of the
 residues but also 
in kinetic energy of the particles emitted to the gas. The mass of the gas 
is shown in the
right hand panel
of Fig.10
    as a function of the relative energy loss. 
One sees that
the 
mass loss (and correspondingly the energy loss)
to the gas is relatively independent of incident energy for MD
interactions 
(black curves), while there is a larger difference for the MI interactions 
(grey/red curves).
Since
the energy lost into the gas is not necessarily closely related to the 
interaction time, this 
represents a correction to the correlation between interaction time and
energy loss, which is 
discussed below.

\begin{figure} [ht]
\unitlength1cm
\begin{center}
\epsfig{file=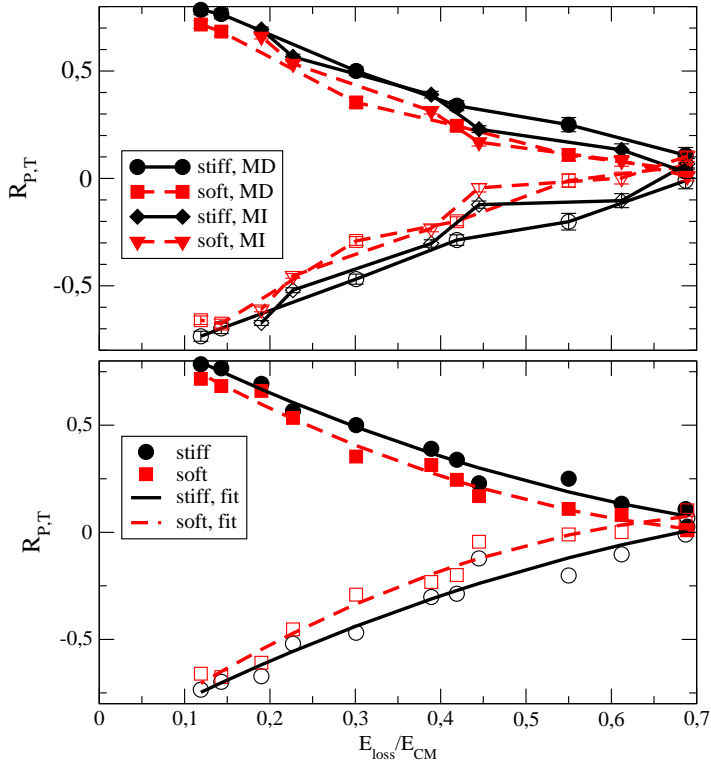,width=10.0cm,angle=-90}
\vspace{0.5cm}
\caption{Imbalance ratio as a function of relative energy loss. 
Upper panel: Separately for 
stiff (solid) and soft (dashed) iso-EoS, and for MD 
(circles and squares) and MI 
(diamonds and triangles) interactions, in the projectile region (full symbols)
 and the target region 
(open symbols).
Lower panel: Quadratic fit to all points for the stiff (solid), resp.
 soft (dashed) 
iso-EoS.
}
\end{center}
\label{imb_eloss}
\end{figure}

In Fig.11  we present the imbalance ratio as a function of 
the relative 
energy loss. In the upper panel we separate the results for MD and MI 
interactions and stiff and 
soft iso-EoS, but in each case the results for 35 and 50 MeV  are  collected 
together and connected by
 lines. In contrast to Fig.9 we now see that the points for the 
different incident 
energies approximately fall on one line, however with considerable scatter 
around it. The 
deviations from a smooth behaviour are larger for the MI interactions 
(diamonds and triangles). This can 
be traced back to the observation, made above, that in this case the 
pre-equilibrium emission, 
i.e. the energy loss to the gas, is rather different for the two 
beam energies, 
while this is not 
so much the case for the MD interaction. Considering that the pre-equilibrium 
emission is not a 
measure of the interaction time, this introduces a correction to the values 
obtained for 35 MeV relative 
to those at 50 MeV, which tends to make the dependence more smooth. We do not 
attempt, however, 
to make this correction quantitative, because this would amount to separating 
the gas particles 
into pre-equilibrium and evaporation components.

It is seen in Fig.11 that the curves for the 
{\it asy-soft} EoS (dashed) are 
generally lower in the projectile region
 (and oppositely for the target region), i.e. show 
more equilibration, that those for the {\it asy-stiff} EoS. In order 
to emphasize 
this trend we have, in 
the lower panel of the figure, collected together  all  the values for the
stiff (circles) and 
the soft (squares) iso-EoS, and fitted them by a quadratic curve. 
It is seen that this fit 
gives a good representation of the trend of the results.

The difference between the curves for the stiff and soft iso-EOS in the 
lower panel then isolates
 the influence of the iso-EoS from kinematical effects 
associated with the 
interaction time. It is seen,
 that there is a systematic effect of the symmetry energy of the order 
of about 20 percent, 
which should be measurable. The correlation suggested in Fig.11 
should represent 
a general feature of isospin diffusion, and it would be of great 
interest to verify 
experimentally.
As discussed above, this kind of analysis would help also in the
comparison with the results of other theoretical models.

On the other hand, it is also seen that the effect is not extremely large, 
 in view  also of the 
substantial scatter of the points around the curves. Thus it is still of 
interest to investigate 
other observables, which might show more sensitivity to the iso-EoS. This 
will be done in the next 
 section  with respect to ternary events.

\section{Ternary  Reactions }

\subsection{Characterization of  Intermediate Mass Fragments }

As discussed above, and seen in the example of Fig.3, there are peripheral
events in
which a third intermediate mass fragment (IMF) appears. The term ``neck
fragmentation'' has been
coined for such events, see \cite{MDT_WCI06} and refs. therein.
Related isospin transport studies have been performed theoretically 
for
symmetric reactions
\cite{isotr05} and theoretically and experimentally for asymmetric collisions
\cite{baranNPA730,chim05}.
The asymmetry and the kinematical properties of the IMF were shown to carry
additional
information on the symmetry energy. Here we investigate this information 
more in detail in the 
system under study.

In such reactions the asymmetry of the IMF is of particular interest. In 
symmetric systems it 
is a result of isospin migration, due to density gradients between the 
residues and the dilute 
neck. In mixed systems this effect is in competition with the isospin 
diffusion discussed in the 
last section due to concentration gradients. While the study of this 
competition may yield 
further information on the time scales of these different processes, the
 more pure case will 
be the study of the symmetric systems.

\begin{figure}
\unitlength1cm
\begin{center}
\epsfig{file=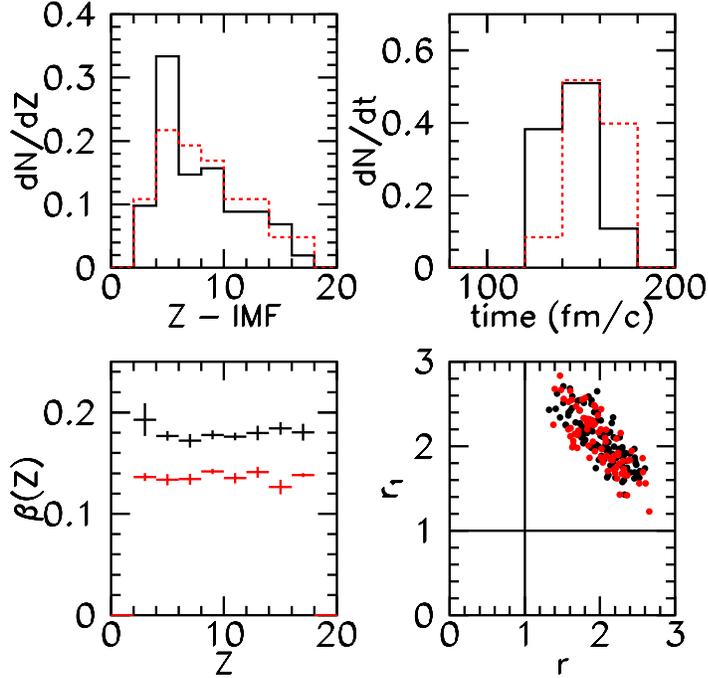,width=10cm}
\caption{Properties of Intermediate Mass Fragments (IMF) in the 
reaction $^{124}Sn+^{112}Sn$ at 
50  $AMeV$  and $b=6fm$ for the stiff (solid lines and black symbols) and
soft (dotted lines, grey/red
symbols) iso-EoS: (upper left) charge distribution; (upper right) time of 
appearance of IMF; 
(lower left) asymmetry distribution of IMF's; (lower right) velocity 
correlation of IMF with 
respect to PLF ($r$) and TLF ($r1$), where $r$ is the ratio of the
relative IMF-residue velocity 
to the Viola systematics velocity.}
\end{center}
\label{IMF_prop}
\end{figure}

We start with characterizing the origin of the IMF's. As an example 
in Fig.12 we 
display various properties of IMF's from the mixed 
reaction $^{124}Sn+^{112}Sn$ at 50  $AMeV$ 
and an impact parameter of $b=6fm$ for the stiff (solid histogram,
black symbols) and soft 
(dotted,  grey/red ) iso-EoS. In the upper left panel the charge 
distribution is
displayed, exhibiting 
a broad distribution  around charge $Z\approx 8$, which is not very
different for different iso-EoS's.
The upper right panel 
shows the times at which the third fragments are identified
 in our method. It is seen that
they appear rather 
late and in a rather short time span. 
The lower left panel 
displays the asymmetry of the IMF's as a function of their charge 
(later we will only show the average asymmetry). This asymmetry 
distribution is rather 
flat; it is, however, substantially different depending on the iso-EoS 
\cite{baranNPA703}. 
Finally in the lower right panel we show velocity correlations between 
the fragments and the residues, as it was suggested in ref. \cite{baranNPA730} 
(so-called Wilczynski-2 plot). The quantity $r$ is the ratio of
the relative velocity of the IMF and the PLF to the Viola systematics
 velocity, which corresponds to 
fission-like events, 
and $r1$ is the same quantity with respect to the TLF. A value of 1 
signifies a late, 
statistical origin of the IMF, while values substantially different from 
one indicate an IMF 
of dynamical origin, which is clearly the case here. Altogether the 
information in this 
figure then suggests that the IMF's come from a uniform source, which 
can be identified with 
the neck region.

\subsection{Isospin Dynamics with Fragment Production}

We have looked at the asymmetries of the various reactions parts (gas, 
residues and exchanged 
particles, like in  Figs.5, 6 and 7) 
also for the 
ternary events. It could be expected that these are different  from those 
of binary events, because some 
of the isospin is 
carried away by the IMF. However, we find that the differences are very 
small and thus we do not 
show these figures separately. The relatively small mass of the IMF's thus 
does not substantially 
influence the asymmetry of the main reaction products.

\begin{figure}
\begin{center}
\epsfig{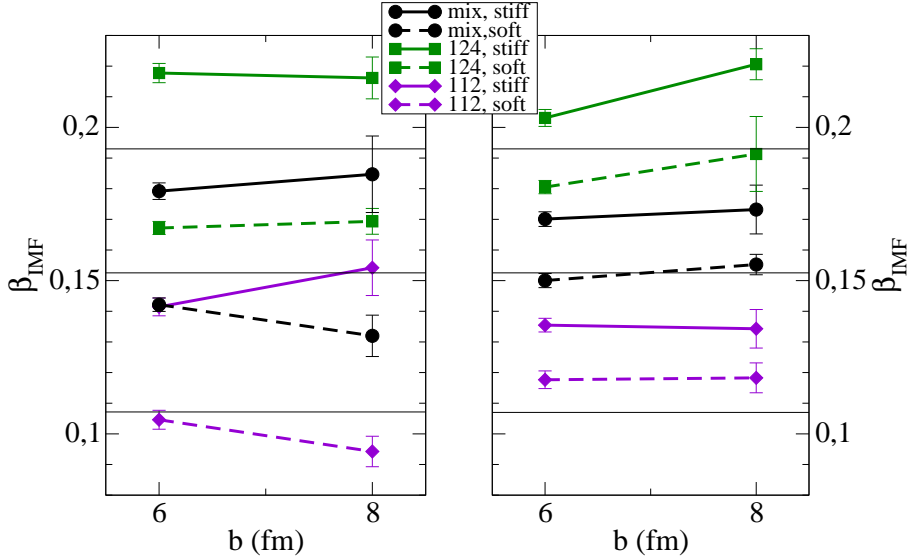}
\caption{Asymmetries of IMF's in ternary $Sn+Sn$ reactions at 50 AMeV 
as a function of impact 
parameter for MD (left panel) and MI (right panel) interactions: symmetric 
124+124 
(squares), symmetric 112+112 (diamonds), mixed 124+112 (circles); stiff 
iso-EoS (solid lines), soft iso-EoS (dashed lines). Horizontal thin 
lines denote the asymmetries 
of $^{124}Sn$, mixed  reaction  and $^{112}Sn$, respectively.
}
\end{center}
\label{asymm_IMF}
\end{figure}

A quantity which is very sensitive to the iso-EoS is the asymmetry 
of the IMF. It is 
shown in Fig.13 at 
50 $AMeV$  for the MD (left panel) and MI (right panel)
 interactions, 
separately for the symmetric and the mixed reactions. The asymmetry of the 
IMF is larger, i.e. 
the IMF is more n-rich, for the stiff  relative to the soft iso-EoS,
since the former  exhibits a larger
isospin migration due to 
the larger slope of the symmetry energy below saturation. This is clearly 
the case for the 
symmetric reactions, but it is also true for the mixed reactions, where 
there is a competition with
isospin diffusion, which depends on the value of the symmetry 
energy and it
is larger for the 
soft iso-EoS. Our result then shows that the isospin migration is the 
dominating 
effect for the 
asymmetry of the neck fragments. 
 
One may now think to investigate the imbalance ratio for the asymmetry of
the IMF, i.e. the ratio
in Eq.(\ref{imb_rat}) where the quantity $x$ represents the corresponding 
 average asymmetry of the IMF's independent of their rapidity. 
As was already seen in ref. \cite{baranNPA703}, this quantity is always 
found to be close to  zero.  This means that the asymmetry of the IMF in
the mixed reaction is the average of that in the symmetric reactions. This 
is expected around mid-rapidity for symmetry reasons, although the
IMF-asymmetries for all the systems, $HH$, $LL$ and $HL$, are different 
from the initial 
    values.
 This is then approximately still true for all the neck fragments. 
In fact we  note that the IMF asymmetries for all
systems are  changing in the same direction, increasing with the
stiffness of the
symmetry term.  This is a non-trivial condition for the  almost zero imbalance
ratio  of the neck fragments.

However, a very striking feature seen in Fig.13  is the fact that the 
difference 
between the  IMF-asymmetry for the stiff and the soft iso-EoS's is large, 
and in addition, that  it  is
much larger in the case of MD interactions relative to MI interactions
 (difference between solid and 
dashed lines in the two panels, for each colliding system). The large 
sensitivity of the IMF 
asymmetry to the iso-EOS was 
already noted earlier \cite{baranNPA703} for MI interactions, showing 
the effectiveness of isospin
migration for  a  stiff iso-EoS.

An interesting point which deserves a deeper interpretation is that the
iso-EoS effect appears 
even more prominent for MD interactions.
 We see that $\beta_{IMF}$ increases for the  {\it asy-stiff} case
(solid lines)
and decreases for the  {\it asy-soft}  one (dashed lines) for 
MD relative to MI 
interactions. Therefore 
the differences in the predicted IMF-asymmetries  for different iso-EoS's  are
systematically
enhanced. This can be traced back to  the following effect  of MD interactions:
 the reaction  is faster and so we  obtain  a mid-rapidity neck fragment only
in regions with larger  projectile-target  overlaps.  Thus, for a given impact
parameter the IMF is produced in a more compact configuration, i.e. with a
 larger
interface with the spectators \cite{jopa_dott} near normal density, where 
the {\it asy-stiff} EoS is effective with respect to neutron migration
to the neck,
while the {\it asy-soft} is not, due to the low value of the derivative
of the symmetry energy close to normal
density (see Fig.1).
In conclusion, with MD
interactions we see a more prominent iso-EoS effect with a strong 
$n$-enrichment of the
neck fragments only for the {\it asy-stiff} case.

\subsection{Asymmetry  Ratios }

\begin{figure}
\begin{center}
\epsfig{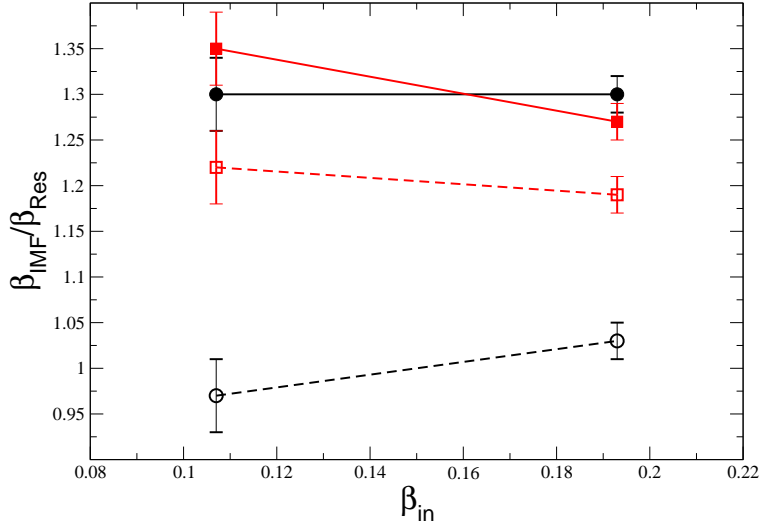}
\caption{ Ratio of asymmetries
of IMF to residues for symmetric $Sn+Sn$
reactions as a function of the initial isospin asymmetry  for
semiperipheral  events ($b=6fm$) at $50 AMeV$ beam energy. 
 Results are shown for MD (circles) and MI (squares) interactions, and 
for stiff 
(solid lines) and soft (dashed lines) iso-EoS's.
}
\end{center}
\label{ratio_IMF_PLF}
\end{figure}

It is of interest to construct an observable, which quantifies these 
effects and which does not 
depend sensitively on measuring the absolute asymmetries, which are 
changed by the secondary 
evaporation. Considering that the asymmetries of the other reaction
 products 
(residues and the gas) is not  equally  sensitive to the iso-EoS
(see  Figs.5
and 6 ; which are for binary events, but, as  stated  above, do not change
noticeably for ternary events), it is attractive to look for ratios of
asymmetries.
In
Fig.14  we show the ratio of the asymmetries of the IMF's to those of the
residues for the symmetric $Sn+Sn$ reactions for stiff (solid) and soft
(dashed) iso-EoS,
and for MD (circles) and MI (squares) interactions. The results
 correspond to $b=6fm$ semiperipheral
events, plotted  here  as a function of the initial isospin asymmetry of the
colliding system.

The ratio between the asymmetry of IMF's and residues can be estimated
on the basis of simple energy balance considerations.
In fact, isospin migration is due to the fact that the neck region
has lower density with respect to the residues and the symmetry 
energy is decreasing with density.
Starting from a residue of mass $A_{res}$ and a neck of mass $A_{IMF}$
of different density  but uniform asymmetry $\beta$, 
we assume that the mass $A$ participating in the isospin exchange
is approximately equal to the mass of the neck, while it is small relative to
the mass of the residue.  This will lead to the asymmetry  
$(\beta + \Delta\beta)$
of the neck, and to a total asymmetry $\beta_{res} = [\beta(A_{res} - A) + 
(\beta - \Delta\beta)A]/A_{res} = \beta - \Delta\beta A/A_{res}$ 
of the residue, with $\Delta\beta$ to be determined by minimization of
the symmetry energy. 
The corresponding variation of the symmetry energy is equal (apart from a 
constant) to:  
\be
\Delta E_{sym} = A_{res} E_{sym}(\rho_R)(\beta-\Delta\beta A/A_{res})^2
+ A E_{sym}(\rho_I)(\beta+\Delta\beta)^2 ,
\ee
where $\rho_R$ and $\rho_I$ are the densities of the residue and
neck regions, respectively.  The minimum of the variation of 
$\Delta E_{sym}$ yields 
\be
\frac{\beta_{IMF}}{\beta_{res}} 
= \frac{E_{sym}(\rho_R)}{E_{sym}(\rho_I)} = 1 +  
\frac{E_{sym}(\rho_R)-E_{sym}(\rho_I)}{E_{sym}(\rho_I)}
\ee
From this simple argument the ratio between the IMF and residue asymmetries 
should
depend only on symmetry energy properties and, in particular, on the 
difference of the 
symmetry energy between the residue and the neck regions, as appropriate 
for isospin migration.  
It should also be larger than one, more so for the {\it asy-stiff} than
for the {\it asy-soft} EoS.

It is seen indeed in Fig.12, that this ratio of IMF over residues asymmetry
is nicely
dependent on the iso-EoS only and not on the initial asymmetry.
This is a
clear feature of the  dominant isospin migration mechanism which depends 
only on
the density gradients.
The ratio is also mostly larger than unity.
In both cases,  MI and MD interactions,
the  {\it asy-stiff} EoS  
is more effective in the n-enrichment of the dilute region where the
IMF's are produced.
Moreover, since for the {\it asy-stiff}  EoS, the derivative of the symmetry
energy with respect to the density is constant, isospin migration
effects should be essentially the same in  MI and MD interactions, as
roughly confirmed by Fig.12.  On the other hand, isospin effects appear
quite reduced for the {\it asy-soft} EoS, especially
for the MD interaction  for the reasons 
already mentioned (reduced slope and larger neutron emission for the 
{\it asy-soft} EoS for the more compact configurations of the MD dynamics).
These behaviours leads to a rather large sensitivity to the iso-EOS for the
 more
realistic MD interactions.
We note that the iso-EoS effect amounts to more than $30\%$. Since we expect
the secondary evaporation to affect in a similar way the asymmetry of IMF's
and residues, we  suggest this observable as very sensitive probe to the
isovector part of the EoS.

\begin{figure}
\begin{center}
\epsfig{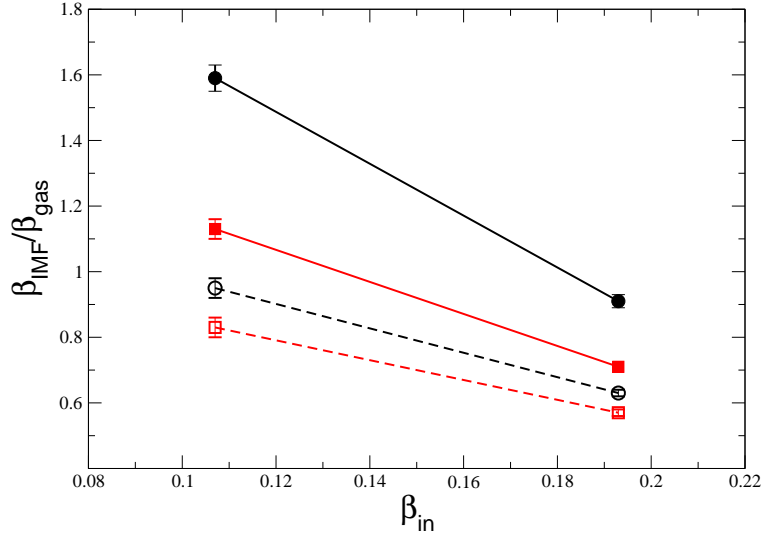}
\caption{ Ratio of asymmetries of IMF to gas for symmetric $Sn+Sn$
reactions as a function of the initial isospin asymmetry. 
Semiperipheral events ($b=6fm$) at $50 AMeV$ beam energy. 
Signatures are as in Fig. 12}
\end{center}
\label{ratio_IMF_gas}
\end{figure}

A similar investigation can be performed with the ratio of the asymmetries
of the IMF to 
 those  of the gas. The corresponding results  
are shown in Fig.15. The mechanism of the gas emission is not simply 
related to the isospin  migration and now the ratios are seen  to be more 
dependent
of the initial asymmetry. The decreasing trend  seen from  $^{112}Sn$ to
$^{124}Sn$ collisions is a consequence of the larger asymmetry of the gas
for the more isospin 
asymmetric systems (see Fig. 5). 
Impressive is
the large value of this ratio for the $^{112}Sn$ collisions and MD 
interaction and 
{\it asy-stiff} EoS, which nicely confirms
the interpretation of a larger proton enrichment and thus smaller asymmetry  
of the gas phase.
Also for this observable a larger sensitivity to the iso-EoS is observed
in the  MD  case.  This reflects the large difference in the predicted IMF
asymmetry between the two  iso-EoS's, as discussed above.

\subsection{Double Ratios}

In order to further reduce the secondary decay effects it has  recently been
proposed to look at double ratios of isospin-dependent quantities, 
 $n$-rich vs. $n$-poor  systems,  e.g. in central collisions for the 
asymmetries
of nucleon \cite{famiano} or 
IMF \cite{MCcentral} emissions. From     Figs.14/15 this double ratio
should  always be around unity for  $\beta_{IMF}/\beta_{Res}$, not much
depending on the initial asymmetry, while it appears of interest for
the  quantity $\beta_{IMF}/\beta{gas}$.
From the same class of semiperipheral events we can extract  Tab.III 
for the Double Ratios:
$$
\frac{{\beta_{IMF}/\beta{gas}}(^{124}Sn+^{124}Sn)}
  {{\beta_{IMF}/\beta{gas}}(^{112}Sn+^{112}Sn)}
$$
We note the almost $20\%$  Iso-EoS  dependence for the more realistic $MD$
interactions.

\noindent

\begin{center}
{\bf Tab. III:}~ $\beta_{IMF}/\beta{gas}$ Double Ratio.
\noindent

\begin{tabular}{c|c|c} \hline
Interaction  &
   MD  & MI \\ \hline
$Asy-soft$  & 0.660  & 0.687  \\ \hline
$Asy-stiff$  & 0.572 & 0.628    \\ \hline

\end{tabular}
\end{center}

\section{Summary and Perspectives}

In this work we have systematically studied isospin transport in heavy
ion collisions in the  Fermi energy domain for symmetric and asymmetric
combinations of $Sn$ isotopes. We have focused on semi-peripheral
dissipative collisions to study both isospin equilibration between
the residues and fragmentation of the neck between the residues,
i.e. to study the isovector part of the EoS for densities below saturation 
density.
The purpose of this investigation has been to analyze in detail the mechanisms
of isospin dynamics and to identify observables, which are sensitive to
the still controversial iso-EoS. We have employed two typically
different iso-EoS's ({\it asy-soft} and {\it asy-stiff}), where one 
expects to see
characteristic signatures since the value and the density-slope of the
symmetry energy determine the effects of isospin migration (due to density
differences) and isospin diffusion (due to isospin concentration differences).

We have clearly shown that the isospin transport is also dependent on
the overall reaction dynamics, mostly ruled by the {\it isoscalar} 
properties of the effective interactions. Thus, it is important to select
observables able to disentangle between isoscalar and isovector
contributions.

In this respect we have thoroughly analyzed the effects of the 
well established (isoscalar)
momentum dependence of the mean field on isospin transport phenomena.
The momentum dependence of the effective forces can affect the
relationship
between isospin transport observables and the symmetry energy
since the reaction dynamics is modified. In particular, the
interaction times are shorter and the composite nuclear systems break
starting from more compact configurations. It is then difficult to
distinguish effects of the iso-EoS and of the momentum-dependence
in the usual analysis as a function of centrality.

For binary dissipative reactions we therefore suggest a study of the
imbalance ratios as a function of the relative energy loss, which is in turn
well correlated to the interaction time. We thus obtain rather well defined
$universal$ curves, including results from different MD interactions
 at different beam energies, which
present, however, a clear dependence on the iso-EoS of the order of 20$\%$.
This kind of analysis should be useful and illuminating in the comparison with
experimental data, as well as in testing the predictions of
different theoretical models. In fact, it allows to disentangle isospin
effects from other model ingredients that may influence the degree
of dissipation reached in the system.

For ternary events, where an Intermediate Mass Fragment (IMF) is formed in 
the mid-rapidity
region (neck fragmentation), the IMF neutron enrichment is very sensitive to
the iso-EoS. We have isolated the isospin
 migration as the dominant mechanism in this process in collisions of the 
symmetric
systems ($^{124}Sn+^{124}Sn~vs.~^{112}Sn+^{112}Sn$). The iso-EoS
dependence is clearly emerging, being stronger for an {\it asy-stiff} 
choice with a
larger slope of the symmetry energy. We have also found that the effect is 
larger
for the more realistic MD dynamics, due to different density regions
which are probed.

All the properties of the reaction products, i.e. residues, gas and IMF's, are
evaluated at the freeze-out time and thus they can be  modified by the 
sequential
decay of the excited products. We therefore propose the investigation
of ``relative'' isospin contents,
i.e. of ratios of the asymmetry $\beta=(N-Z)/A$ of the various products,
which are expected
to be less affected by secondary emissions. A very promising quantity
seems to be the ratio $\beta_{IMF}/\beta_{Res}$ at mid-centrality,
which directly measures the isospin migration. In fact, it is found to be very
sensitive to the iso-EoS, up to above the $30\%$ level for the more
realistic MD interactions.

In conclusion, we suggest new isospin sensitive
observables to study in dissipative HIC at Fermi energies in
order to further constrain
value and slope of the symmetry term at subsaturation densities. 

It should also be noticed that the ratios considered here 
(Imbalance Ratios, $\beta_{IMF}/\beta_{Res}$) do not depend,
at a first level approximation,  
on the
initial asymmetry of the systems considered, but only on the 
properties (value and slope) of the symmetry energy. Hence valuable information
on these properties 
can already be obtained with the available neutron-rich beams.  
However, the use of
radioactive beams 
in this intermediate energy range 
would be certainly important in order to extend the systematics of results
and to better constrain the symmetry energy behaviour.  

\vskip 0.5cm

{\it Acknowledgments}

We warmly thank M.B.Tsang for fruitful discussions.
This work was supported partly by the German Ministry for
Education and Research 
(BMBF), grant 06LM189, and by the
DFG Cluster of Excellence {\it Origin and
Structure of the Universe} (www.universe-cluster.de).
V.B. acknowledges the support of the Romanian Minister for Education and
Research, under the contracts No.CEX-05-D10-02 and CEX-05-D11-03. 
V.B., H.H.W. and M.Z.-P.
are grateful to the LNS-INFN for the
warm hospitality during part of this work.


\end{document}